\documentclass[a4,center,fleqn]{NAR}

\usepackage{gensymb} 
\usepackage[version=4]{mhchem}
\usepackage{ulem} 
\usepackage{anyfontsize}

\copyrightyear{YYYY}
\pubdate{DD MM YYYY}
\pubyear{YYYY}
\jvolume{xx}
\jissue{xx}

\begin{document}

\title{Stacking correlation length in single-stranded DNA}

\author{%
Xavier Viader-Godoy\,$^{1,2}$,
Maria Manosas\,$^{1}$
and Felix Ritort\,$^{1}$%
\footnote{To whom correspondence should be addressed.
E-mail: ritort@ub.edu}}

\address{%
$^{1}$Small Biosystems Lab, Departament de F\'isica de la Mat\`eria Condensada, Facultat de F\'isica, Universitat de Barcelona, Carrer de Martí i Franquès, 1, 08028 Barcelona, Spain.
$^{2}$Dipartimento di Fisica e Astronomia Galileo Galilei, Università degli Studi di Padova, Via Francesco Marzolo, 8, 35131 Padova, Italy.}

\history{%
Received YYYY-MM-DD;
Revised YYYY-MM-DD;
Accepted YYYY-MM-DD}

\maketitle

\begin{abstract}
Base stacking is crucial in nucleic acid stabilization, from DNA duplex hybridization to single-stranded DNA (ssDNA) protein binding. 
While stacking energies are tiny in ssDNA, they are inextricably mixed with hydrogen bonding in DNA base pairing, making their measurement challenging. 
We conduct unzipping experiments with optical tweezers of short poly-purine (dA and alternating dG and dA) sequences of 20-40 bases. We introduce a helix-coil model of the stacking-unstacking transition that includes finite length effects and reproduces the force-extension curves.  Fitting the model to the experimental data, we derive the stacking energy per base, finding the salt-independent value $\Delta G_0^{ST}=0.14(3)$ kcal/mol for poly-dA and $\Delta G_0^{ST}=0.07(3)$ kcal/mol for poly-dGdA. Stacking in these polymeric sequences is predominantly cooperative with a correlation length of $\sim 4$ bases at zero force. The correlation length reaches a maximum of $\sim 10$ and $5$ bases at the stacking-unstacking transition force of $\sim 10$ and $20$ pN for poly-dA and poly-dGdA, respectively.
The salt dependencies of the cooperativity parameter in ssDNA and the energy of DNA hybridization are in agreement, suggesting that double-helix stability is primarily due to stacking. 
Analysis of poly-rA and poly-rC RNA sequences shows a larger stacking stability but a lower stacking correlation length of $\sim 2$ bases.

\end{abstract}

\section{Introduction}

Nucleic acids (NAs) participate in information transfer and regulatory genomic processes that require the readout of the bases. Molecular reactions in NAs involve opening double-stranded (ds) helical structures, converting them into single-stranded forms (ssDNA and ssRNA), making bases accessible to the cell machinery \cite{saenger1984}. Stacking forces are crucial in the hybridization reaction where the two complementary strands form a duplex also stabilized by hydrogen bonding \cite{Holbrook1999,Whitley2017}. 
Base stacking is also essential for understanding allostery and other molecular actions at a distance. In DNA, stacking regulates protein binding during replication and recombination \cite{Kim1995}. Stacking of the nascent RNA chain modulates co-transcriptional RNA folding, whereas for mRNA it impacts co-translational protein folding \cite{Goldman2015,bustamante_yan_2022}.

Besides, stacking in ssDNA promotes the formation of structures \cite{Protozanova2004,Yakovchuk2006,viader2021prx}.
Poly-deoxyadenine (poly-dA) sequences form single and double-stranded helices \cite{Chakraborty2009} stabilized by stacking interactions, similarly to poly-adenine (poly-rA) sequences \cite{Rich1961}. In contrast, poly-deoxycytosine (poly-dC) and poly-deoxyguanine (poly-dG) sequences form complex structures such as i-tetraplexes and G-quadruplexes, relevant for the regulation of gene expression \cite{Gehring1993, Assi2018, Robinson2021}. These polynucleotide structures emerge from the interplay between stacking and non-canonical base pairing. The structural diversity of ssDNA is also relevant for many applications such as DNA origami \cite{Castro2011}, DNA nano switches \cite{Green2019}, synthetic molecular motors \cite{Balzani2000} and immunodetection \cite{Assi2018}. Despite their importance, direct measurements of stacking energies in ssDNA remain scarce. 

\enlargethispage{-65.1pt}

Base pairs in dsDNA form adjacent stacks that stabilize the double helix. Stacking energies in dsDNA have been measured using DNA origami nanotubes manipulated with optical tweezers \cite{Kilchherr2016}, and indirectly through melting experiments \cite{SantaLucia1998} and by mechanically unzipping single DNA molecules \cite{Huguet2010PNAS,huguet2017NAR}.
The energies of the ten different combinations of stacks in the nearest-neighbor (NN) model have been determined, finding values in the range of $1-3$ kcal/mol \cite{Gray1970,Licinio2007}. Various studies indicate that stacking is the main contribution to the free energy of hybridization \cite{Protozanova2004,Yakovchuk2006,Bosaeus2012,Feng2019,Punnoose2023}. However, these measurements do not permit us to infer the much lower stacking energies of ssDNA, approximately $\sim0.1$ kcal/mol \cite{saenger1984}.
Stacking of ssDNA has been measured with several techniques: calorimetry \cite{ramprakash2008biopol}, nuclear magnetic resonance \cite{Isaksson2004}, X-ray diffraction \cite{plumridge2017}, single-molecule fluorescence \cite{Aalberts2003}, atomic force microscopy \cite{Ke2007}, and magnetic tweezers \cite{mcintosh2014}. Measurements of force-extension curves with magnetic tweezers on purine-rich sequences \cite{mcintosh2014} show a cooperative stacking-unstacking (S-U) transition around $20$ pN. Most NA studies have focused on homopolymeric sequences containing purines or pyrimidines. Poly-dA shows the largest level of stacking \cite{Ke2007}, whereas poly-uracil (poly-U), poly-deoxythymidine (poly-dT), and mixed poly-pyrimidine (poly-pyr) sequences do not show stacking \cite{Ke2007,Aalberts2003,mcintosh2014}. On the other hand, studies with short DNA oligonucleotides of mixed sequences demonstrate that the minimal purine-rich motif for stacking must contain at least four bases \cite{ramprakash2008biopol}.

Here, we investigate stacking in ssDNA by measuring the force-extension curves (FECs) of short poly-purine sequences of varying lengths using optical tweezers. The FECs exhibit a shoulder at a given force, where the convexity of the FEC changes, a feature of the stacking-unstacking transition. We introduce a helix-coil model for stacking that reproduces the experimental FECs of poly-dA and poly-dGdA and the observed finite length effects. The model contains two energy parameters: the stacking energy per base $\epsilon_{\textrm ST}$ and the cooperativity of stacking between neighboring bases $\gamma_{\textrm ST}$. The sensitivity of the force data and the model's features permit us to accurately derive the stacking free energies and correlation length at different salt conditions. Notably, we find a maximum in the correlation length at a force value directly related to the energy parameters of the model, $\epsilon_{\textrm ST}$ and $\gamma_{\textrm ST}$.
Finally, we further validate the model by analyzing previous results on different ssDNA \cite{mcintosh2014} and ssRNA \cite{Seol2007prl} sequences. Our results show that $\gamma_{\textrm ST}$ is systematically larger than $\epsilon_{\textrm ST}$, indicating that stacking cooperativity is the primary source of stabilization in nucleic acids.

\section{MATERIALS AND METHODS}

\begin{figure*}[t]
\begin{center}
\includegraphics{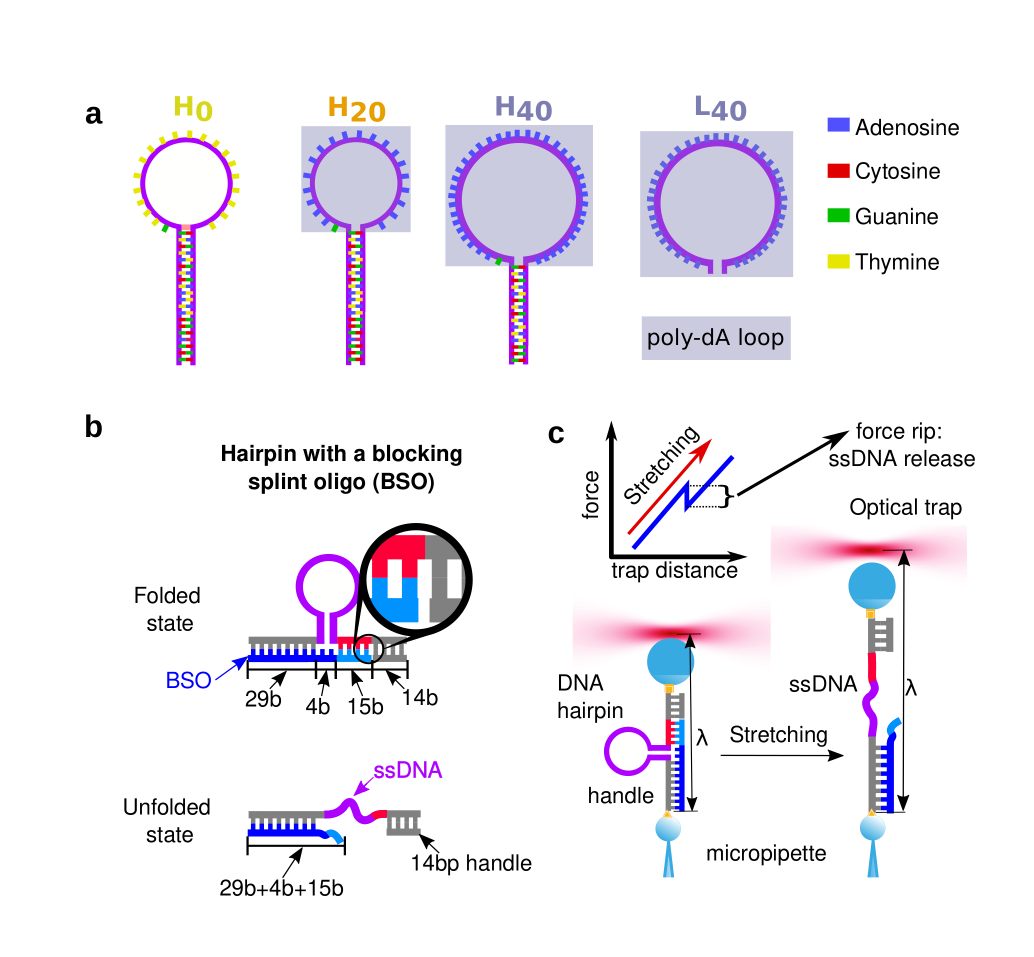}
\end{center}
\caption{{\bf Hairpin sequences, molecular construct and experimental setup.} {\bf a.} Schematic depictions of the four hairpins studied. Grey boxes show the poly-dA regions of the sequence. Hairpins are named by the number of purines in the loop: $H_{\textrm{0}}$ (20bp stem and 20b dT-loop); $H_{\textrm{20}}$ (20bp stem and 20b purine-loop -1dG, 19dA-); $H_{\textrm{40}}$ (15bp stem and 40b purine-loop -1dG, 39dA-). $L_{\textrm{40}}$ consists of a 40dA loop without stem. {\bf b.} Scheme of the folded (top) and unfolded (bottom) states of the hairpins with the 48b blocking splint oligonucleotide (BSO). {\bf c.} Sketch of the optical tweezers setup. Left: a DNA hairpin is attached to two beads using specific linkages, one is held by a micropipette and the other by the optical trap. Right: the trap distance $\lambda$ is moved away from the micropipette, while the applied force on the DNA hairpin increases, until it unfolds, and a force rip is observed in the force-distance curve (top left).}\label{fig:1}
\end{figure*}

\subsection*{Instrument and experiments}
Experiments were performed using the miniTweezers setup \cite{Smith2003}, which consists on two counter-propagating laser beams (P=200 mW, $\lambda$=845 nm) that follow a symmetrical optical path and are focused to create a single optical trap in a microfluidics chamber \cite{gieseler2020optical}. The molecular construct was tethered between polystyrene beads coated with Streptavidin (2.0 $\mu$m \textit{Kisker Biotech}) or anti-Digoxigenin (3.0 $\mu$m \textit{Kisker Biotech}), keeping the former fixed via air suction to a micropipette. The latter was trapped by the optical trap whose exerted force and position were measured by using Position Sensitive Detectors (PSDs) \cite{Zaltron2020EPJP}. The instrument has a 0.1 pN and 1 nm resolution at a 1 kHz acquisition rate. Beads were attached to the DNA construct by labelling each molecular end with Biotin or Digoxigenin.
The DNA hairpins used in this work were synthesized by following a similar procedure to what was done in Ref.~\cite{Landuzzi2020}, annealing three oligonucleotides (Supplementary Sec.S1).
The first and longer one contains the hairpin region and the flanking 29b of the handles (with a Biotin in its 5' end, \textit{Merck Sigma-Aldrich}), and a tail of digoxigenins is added via Terminal Transferase enzyme (\textit{Merck Sigma-Aldrich}). The second oligo is a short 14b segment complementary to the second handle, while the third oligo is fully complementary to the first handle and to the 15b of the second one, leaving a 4b spacer between them.
The experiments were all performed at 25\degree C. Those with varying NaCl concentrations (10, 50, 100, 500 and 1000 mM) contained also (10 mM Tris-HCl pH7.5, EDTA 1 mM, 0.01\% sodium azide), while the experiments with \ce{MgCl2} were performed at 10 mM Tris-HCl pH7.5, 0.01\%\ce{NaN3} (sodium azide) and 10 mM \ce{MgCl2}.

\subsection*{Extension determination}

To obtain the ssDNA extension of the unfolded hairpin, we employ the two-branches method \cite{Zhang2009science}, which is based on analyzing the hairpin's pulling curves. Pulling experiments, where the force is cyclically increased and decreased to unzip and re-zip the hairpin, were carried out using the blocking splint oligo (BSO) method \cite{viader2021ijms} to increase the hairpin unfolding forces. Briefly, a 48-base oligonucleotide is hybridized to the ssDNA handles at the two flanking sides of the hairpin (Fig.~\ref{fig:1}b): 29b hybridize to the left handle (grey-blue duplex in Fig.~\ref{fig:1}b, top); 15b hybridize to the right handle (cyan-red duplex in Fig.~\ref{fig:1}b, top); a spacer of 4b connects both sides to accommodate the diameter of the DNA hairpin stem. When the force increases above $\sim$ 40 pN, the shorter 15b right handle unbinds and the hairpin unfolds keeping the longer 29b left handle hybridized (Fig.~\ref{fig:1}c, right). Upon releasing the force, the hairpin refolds, and the 15b right oligo hybridizes again (cyan segment in Fig.~\ref{fig:1}c, left). The higher stability of dsDNA to shearing increases the hairpin unfolding force from $\sim$15 to $\sim$40pN, allowing us to obtain precise ssDNA FECs for short molecules (See Supp. Sec. S7) in a wide range of forces ($5\leq f\leq45$ pN).
The hairpin force-distance curve (FDC) is divided into two branches: the folded (F), where the hairpin is folded, and the unfolded (U), where the hairpin is unfolded, Fig. ~\ref{fig:2}a. The molecular extension of the ssDNA at a given force, $X_{\rm ssDNA}(f)$, can be obtained by subtracting the relative trap position ($\lambda$) of the U ($\lambda_U$ ) and F ($\lambda_F$) branches. As shown in Fig.~\ref{fig:2}a, the optical trap position at the F branch is given by $\lambda_F=x_{h1}^{\rm 29bp}(f)+x_d(f)+x_{h2}^{\rm 29bp}(f)+x_t(f)$, with $x_{h1(h2)}^{\rm 29bp}$, $x_d$ and $x_t$ being the extension of dsDNA handles 1(2), the oriented hairpin diameter and the bead position in the trap respectively. Whereas the optical trap position at the U branch is given by $\lambda_U=x_{h1}^{\rm 29bp}(f)+X_{\rm ssDNA}(f)+x_{h2}^{\rm 14bp}(f)+x_t(f)$. Therefore:
\begin{equation}
 X_{\rm ssDNA}(f)=\lambda_U(f)-\lambda_F(f)+x_{\rm d}(f)+x_{\rm dsDNA}^{\rm 15bp}(f),
 \label{eq:ssDNA_extension}
\end{equation}
where the oriented hairpin diameter $x_{\rm d}(f)$ is modeled as a Freely-Jointed Chain with a single monomer of 2nm length \cite{alemany2014biopol}; whereas $x_{\rm dsDNA}^{\rm 15bp}(f)$ is the extension of the 15bp segment of dsDNA (Supplementary Sec.S2). 

To compare the elastic behavior of the different ssDNA loop sequences and lengths, we use the re-scaled extension per base in the loop $x_b$, as shown in Figs.~\ref{fig:3}c, \ref{fig:5}a and \ref{fig:6}a for poly-dA and poly-dGdA loops. It is computed by subtracting the elastic contribution of the bases not belonging to the loop (the hairpin stem + 15b of the right handle) to the measured extension $X_{\rm ssDNA}$:

\begin{equation}
 x_b=\frac{X_{\rm ssDNA}-\left(N-N_{\rm loop}\right)x_U}{N_{\rm loop}},
 \label{eq:ssDNA_extension_resc}
\end{equation}

where $N_{\rm loop}$ is the number of bases of each hairpin loop and $x_U$ is the extension of a single unstacked base as given by the TC model (see Results Section). The assumption that the bases outside the loop are unstacked is justified by the analysis presented in Fig.~\ref{fig:2}. For the analysis of the ssDNA and ssRNA data from previous works \cite{mcintosh2014,Seol2007prl}, presented in Figs.~\ref{fig:5}a,c and \ref{fig:6}c, the total number of bases $N$ for each studied molecule (poly-dA, mixed, poly-pyrimidine, poly-U, poly-rA and poly-rC) is obtained by imposing the TC model elasticity at forces $30\leq f\leq50$ pN. Their re-scaled extension is computed as $x_b=x_{\textrm{ssDNA}}/N$. For the mixed sequence (Fig.~\ref{fig:5}c), the same approach described by Eq.~\ref{eq:ssDNA_extension_resc} is followed, subtracting the extension of the bases that do not belong to any poly-purine domain, i.e. that are not bold in the sequence shown in the caption of Figure~\ref{fig:5}.

\subsection*{Helix-coil stacking (ST)-model}

To analyze the stacking-unstacking transition observed in the experiments, we use a helix-coil type model denoted as the stacking (ST) model. The ssDNA molecule is modeled as a polymer chain of $N$ {\it stackable} bases, that can form stacked (S, blue) and unstacked (U, yellow) domains, Fig.~\ref{fig:4}a. The ST model can be mapped onto a one-dimensional Ising chain where each base $i=1,2,..,N$ is represented by a binary variable $\sigma_{i}$. Bases in an S-domain ($\sigma_i=1$) and those in a U-domain ($\sigma_i=-1$) exhibit different elastic responses. 

Two energy parameters define the model (Fig.~\ref{fig:4}a): the (positive) energy gain per stacked base, $\epsilon_{\textrm{ST}}$; and a cooperativity parameter between adjacent domains, $\gamma_{\textrm{ST}}$. The elastic response of U-domains is modeled using the Thick-Chain (TC) model (Supp. Sec. S2), which accounts for steric effects due to the high flexibility of longer U-domains, especially at low salt concentrations. The extension per U-base in the TC model, $x_U(f)$, is described by parameters $l_U$, $a$, and $\Delta$, as previously explained. 
In contrast, the high rigidity of the shorter S-domains is well described by the semiflexible WLC model (Supp. Sec. S2). The extension per base of S-domains, $x_S(f)$ depends on the persistence length, $p_S$, and the contour length per base, $l_S$ and is obtained by inverting the interpolation formula of the model proposed in Ref.~\cite{marko1995stretching}. The total extension $X$ of the chain is $X=N_U\,x_U+N_S\,x_S$, where $N_{U}=\sum_{i=1,N}(1-\sigma_i)/2$ and $N_{S}=\sum_{i=1,N}(1+\sigma_i)/2$ are the total number of unstacked and stacked bases, respectively. The normalized extension per base is $x_{b}=X/N$ with $N=N_S+N_U$ being the total number of bases. Upon increasing the force $f$, the longer U-domains become energetically favored ($N_{U}$ increases while $N_{S}$ decreases), as illustrated in \ref{fig:4}a. The energy function of the ST-model reads:
\begin{eqnarray}
 E(\{\sigma_i\}) = -N_{\textrm{S}}(\{\sigma_i\})\,\Big(\epsilon_{\rm ST}+\int_0^{f}x_{\textrm{S}}(f')df'\Big)- \nonumber \\ 
 -N_{\textrm{U}}(\{\sigma_i\})\int_0^{f}x_{\textrm{U}}(f')df'-\gamma_{\textrm{ST}}\sum_{i=0}^{N+1}\sigma_i\sigma_{i+1}.
\label{eq:hamiltonian_1}
\end{eqnarray}

The integrals in \eqref{eq:hamiltonian_1} are the stretching energy contributions per base in the S and U domains. Fixed (Neumann-Neumann) boundary conditions are imposed at the ends, with $\sigma_0=\sigma_{N+1}=-1$ for the {\it non-stackable} bases outside the poly-dA (or poly-dGdA) region. We have solved the free energy $G(f)$ of the ST-model of Eq.~\eqref{eq:hamiltonian_1} and derived the FECs using $X(f)=-\frac{\partial G}{\partial f}$. Detailed calculations are provided in Supp. Sec. 5. 

For $N=\infty$, the free energy difference between the relaxed ssDNA at zero force and the fully unstacked state (given by Eq.~\eqref{eq:hamiltonian_1} with $\sigma_i=-1,\forall i$) is (Supp. Sec. 5.1):
\begin{eqnarray}
 \Delta G_0^{\textrm ST}=\frac{\epsilon_{\textrm{ST}}}{2}+\frac{1}{\beta}\log\Bigl[\cosh\left(\beta\frac{\epsilon_{\textrm{ST}}}{2}\right)+ \nonumber\\
 +\sqrt{e^{-4\beta\gamma_{\textrm{ST}}}+\sinh^{2}\left(\beta\frac{\epsilon_{\textrm{ST}}}{2}\right)}\Bigr],
\label{eq:free_energy}
\end{eqnarray}
where $\beta=1/k_{\textrm{B}}T$, with $k_{\textrm{B}}$ being Boltzmann's constant and $T$ the temperature. 

The ST model also allows computing the stacking correlation length, $\xi_{\textrm ST}$. Starting from any base of the chain in a state $\sigma_i$, $\xi_{\textrm ST}$ is defined as the distance in nucleotides where correlations of $\sigma$ decay due to thermal fluctuations, i.e. $\left< \sigma_i\sigma_j\right>\sim \exp{\left((j-i)/\xi_{\textrm ST}\right)}$. In the long chain limit ($N=\infty$) the stacking correlation length $\xi_{\textrm ST}$ is computed as (Supp. Sec. 5.1, Eq. S16): 
\begin{equation}
\xi_{\textrm ST} =-\left[\log\left(\frac{\cosh\left(\beta A\right)-\sqrt{e^{-4\beta \gamma_{\textrm{ST}}}+\sinh^2\left(\beta A\right)}}{\cosh\left(\beta A\right)+\sqrt{e^{-4\beta \gamma_{\textrm{ST}}}+\sinh^2\left(\beta A\right)}}\right)\right]^{-1},
 \label{eq:correlation5}
\end{equation}
where $A$ is defined as: 
\begin{equation}
A= \frac{\epsilon_{\textrm{ST}}}{2}-\frac{1}{2}\int_0^{f}\Delta x(f')df', 
\label{eq:termsA}
\end{equation}
with $\Delta x=x_U-x_S$.
As shown in Eq.~\eqref{eq:correlation5}, $\xi_{\textrm ST}$ depends on the value of the force, and it is maximum at the force where FECs exhibit a force-shoulder, indicative of a first-order phase transition.

\section{RESULTS AND DISCUSSION}

We unzip DNA hairpins with optical tweezers and measure {the elastic response of poly-dA tracks using DNA hairpin constructs with poly-dA loops of 20 and 40 bases (b) and stems of 15 and 20 bps ($H_{\textrm{20}}$ and $H_{\textrm{40}}$ in Fig.~\ref{fig:1}a). As a reference, we also study a hairpin sequence with a poly-deoxythymidine (poly-dT) loop of 20 bases ($H_{\textrm{0}}$) and a 40b poly-dA loop without stem ($L_{\textrm{40}}$), Fig.~\ref{fig:1}a. Stacking is primarily contained in the loop because strands in the stem only contain segments of very short, typically less than four, consecutive purines (see below, Supp. Fig. S1 and Supp. Table S1) which will be denoted as {\it non-stackable} sequences. As a comparison we also investigate the elastic response of DNA hairpins with loops of alternating deoxyadenine and deoxyguanine (poly-dGdA) using similar constructs (Fig. ~\ref{fig:6}a). Molecules were pulled from their ends using specifically designed DNA handles and a Blocking Splint Oligonucleotide (BSO) of 48b that links the two handles \cite{Manosas2017,viader2021ijms} (Fig.~\ref{fig:1}b and Methods). The higher stability to shearing of the dsDNA handle than to the unfolding of the hairpins allows us to obtain the ssDNA molecular extension of the purine loops in a broad range of forces ($5\leq f\leq45$ pN) (Fig.~\ref{fig:1}c).

\subsection{Unstacked elasticity}

To measure the FECs of the purine stretches (poly-dA and poly-dGdA) in the loops, we must subtract the contribution of the stem and the BSO to the measured molecular extension (Methods). To derive the elastic response of the {\it non-stackable} ssDNA sequence in the stem, we have studied the reference hairpin $H_\textrm{0}$. Figure~\ref{fig:2}a shows the force $f$ versus relative trap position $\lambda$, the so-called force-distance curve (FDC), in one pulling cycle. The folded hairpin is pulled starting from $\sim$5pN at a constant loading rate (Fig.~\ref{fig:2}a, green line) until a force is reached ($\sim 40$ pN) where the BSO partially detaches (blue oligonucleotide in Fig.~\ref{fig:1}b, bottom), the hairpin unfolds, and a force rip is observed (Fig.~\ref{fig:2}a, top arrow). Upon reaching $\sim 50$ pN the process is reversed, and the force decreases at the same unloading rate (Fig.~\ref{fig:2}a, blue line). 

At $\sim 4$ pN, a force jump event is observed (Fig.~\ref{fig:2}a, bottom arrow), where the hairpin refolds, and the BSO rebinds to the right handle (Fig.~\ref{fig:1}b, top). The difference in extension $\Delta\lambda=\lambda_U-\lambda_F$ between the folded and unfolded branches at a certain force (Fig.~\ref{fig:2}a) gives the FEC of the released ssDNA from the hairpin (60 bases) plus the 15 bases released by the BSO (light blue in Fig.~\ref{fig:1}b, and Methods). Figure~\ref{fig:2}b shows the FEC of the total released ssDNA (75 bases) in 1M \ce{NaCl} (brown circles) and 10mM \ce{MgCl2} (yellow circles). FECs at the two salt conditions are compatible, in agreement with the 1:100 salt rule of thumb, which states that the screening effect at a given concentration in magnesium equals that at $100\times$ concentration in sodium \cite{bosco2014nar,Rissone2022}. Results agree with the FECs measured without the BSO at low forces, $4<f<15$ pN. They are also consistent with previous measurements in a poly-pyrimidine sequence \cite{mcintosh2014} (Supp. Fig. S2) confirming that ssDNA from $H_\textrm{0}$ is fully unstacked.

\begin{figure*}[t]
\begin{center}
\includegraphics{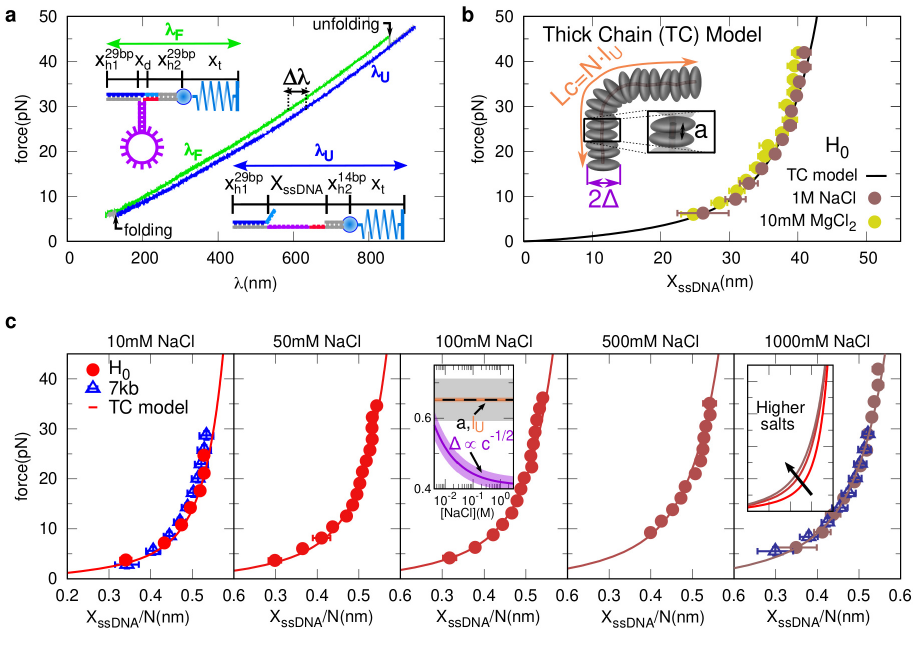}
\end{center}
\caption{{\bf Unstacked ssDNA elasticity.} {\bf a.} A typical force-distance curve (FDC) for $H_{\textrm 0}$. In green (blue) are shown the unfolding (folding) branches, with $\lambda_F$ ($\lambda_U$) being the trap position (trap-pipette distance). The elastic contributions to the trap position $\lambda$ at the F(U) branches are schematically depicted on top (bottom). Arrows indicate the jump in force when the molecule unfolds or folds, changing from one branch to another. These forces dictate the limits for applicability of the two branches method. {\bf b.} Force-extension curve (FEC) of $H_{\textrm 0}$ for 10 mM \ce{MgCl2} (yellow dots) and 1M \ce{NaCl} (brown dots). Black lines show the fit of the TC model to the NaCl data. The inset shows a schematic depiction of the Thick Chain (TC) model, with its three parameters: the disk radius $\Delta$, the spacing $a$, and the total contour length, $L_c$. {\bf c.} FECs per base $x_{\rm ssDNA} /N$, for $H_{\textrm 0}$ (red circles) and 7kbp hairpin [(blue triangles, Ref.~\cite{viader2021prx}] for different \ce{NaCl} concentration, with their fits to the TC model (continuous lines). The inset of the central panel shows the salt dependence of the fitting TC model parameters. Shadowed areas are the statistical errors obtained by bootstrapping (N=500). The right inset shows how the theoretical FECs change with salt concentration. The error bars are the standard errors of the molecules studied at each condition (Supp. Fig. S3, Supp. Table S6).}
\label{fig:2}
\end{figure*}

Figure \ref{fig:2}c shows FECs at 10, 50, 100, 500, and 1000mM NaCl (circles) plotted against the molecular extension per base $x_{\rm ssDNA}/N$. As a comparison, the results for a 7.2kb ssDNA in glyoxal \cite{viader2021prx}, which prevents secondary structure formation, are also shown (triangles in the first and last panels of Fig.~\ref{fig:2}c). The 7.2kb results agree with those of $H_\textrm{0}$, showing that heterogeneous ssDNA sequences lacking many consecutive purines exhibit the same ideal elastic response and can be considered {\it non-stackable} sequences. 

FECs of unstacked ssDNA can be fitted to the Thick Chain (TC) model \cite{toan2005biophys} (Methods, Supp. Sec. S2 and Supp. Fig. S4) over three decades of salt concentration (continuous lines in Figs.~\ref{fig:2}b,c). The TC model conceptualizes the ssDNA as a necklace of contour length $L_c$ consisting of oblate disks of diameter $2\Delta$ and spacing $a$. Disks occupy a finite volume to model steric and electrostatic effects (Fig.~\ref{fig:2}b and methods). Fitting the TC model to the data (Supp. Sec. 3 and Supp. Tables S2-S5) we obtain $l_U=L_c/N=0.652(7)$ nm (contour length per base), $a=0.65(6)$ nm, and a Debye-like salt dependence for the effective radius, $\Delta=0.40(2)+0.0109(13)/\sqrt{C}$, with $C$ the salt concentration in M units. Numbers in parenthesis are the statistical errors in the last digit, obtained from bootstrapping the fitted data points. A similar salt dependence for $\Delta$ has been found for RNA poly-U chains \cite{Toan2006}. The parameters $l_U$ and $a$ are salt independent and compatible with each other (orange and black dashed line, central panel of Fig.~\ref{fig:2}c), showing that one disk in the TC model corresponds to a single base of the ssDNA. The value $l_U\simeq a\simeq 0.65$ nm agrees with the reported crystallographic inter-phosphate distance in ssDNA \cite{camunas2016annurev}. Moreover, half the dsDNA helix diameter ($d_{\rm dsDNA}\sim$2nm) is compatible with the ssDNA radius predicted by the TC model $\Delta=d_{\rm dsDNA}/4\sim 0.5$ nm (magenta line, central panel of Fig.~\ref{fig:2}c). Figure.~\ref{fig:2}c (rightmost panel, inset) shows the fitted FECs at different salt concentrations. The TC model predicts a persistence length, $p_U=-a/\log{\left(1-a^2/(4\Delta^2)\right)}$,with values ranging from 1.3nm (10mM NaCl) to 0.7nm (1M NaCl), consistent with the literature \cite{camunas2016annurev,Bustamante2021}.

\subsection{Poly-dA stacking}

The FECs for all constructs in Fig.~\ref{fig:1}a are shown in Fig.~\ref{fig:3}a, averaged over several molecules at 10mM \ce{MgCl2} (Supp. Fig. S5). If plotted versus the normalized molecular extension $X_{\textrm{ssDNA}}/N$, hairpins $H_{\textrm{20}}$, $H_{\textrm{40}}$ and $L_{\textrm{40}}$ show a shorter extension upon increasing the poly-dA loop size (Fig.~\ref{fig:3}b). For $H_{\textrm{40}}$ and $L_{\textrm{40}}$ a nascent shoulder in the FEC is visible around 15pN, a fingerprint of the unstacking transition \cite{Seol2007prl}. This shoulder appears as a change in the FEC convexity that deviates from the unstacked elastic response, represented by $x_U$. To extract the contribution of the poly-dA bases in the loop from the FECs, we subtract the elastic contribution of the {\it non-stackable} bases of the stem and the 15b of the BSO (Supp. Sec. S4 and Supp. Fig. S6). The FECs for the loops are shown in Fig.~\ref{fig:3}c, with $(HL)_{\textrm{40}}$ being the average of the indistinguishable $H_{\textrm{40}}$ and $L_{\textrm{40}}$ (Supp. Fig. S7). The unstacking transition is now apparent in the FECs, where the shoulder becomes more prominent for larger loop sizes. These finite-size effects demonstrate that the unstacking transition is cooperative. 

\begin{figure}[t]
\begin{center}
\includegraphics{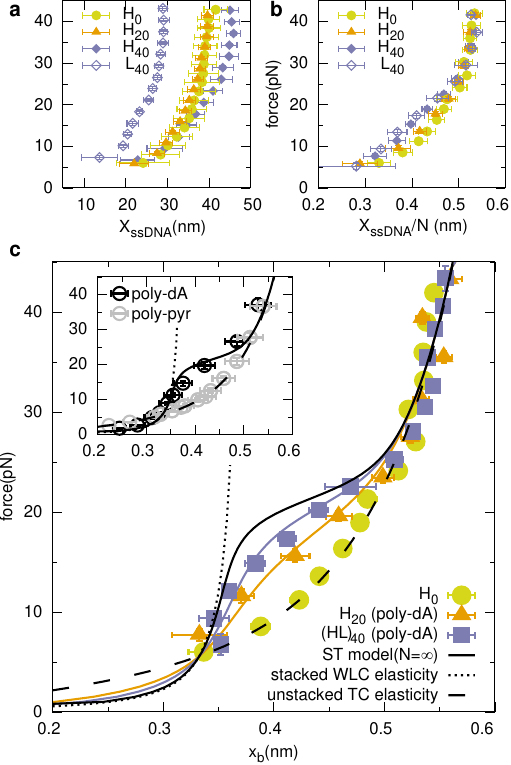}
\end{center}
\caption{{\bf poly-dA stacking} {\bf a.} FECs of $H_{\textrm{0}}$ (yellow circles), $H_{\textrm{20}}$ (orange triangles), $H_{\textrm{40}}$ (blue filled rhombi) and $L_{\textrm{40}}$ (blue empty rhombi). {\bf b.} Re-scaled FECs of the molecules shown in panel a. {\bf c.} Re-scaled FECs for varying lengths: $H_{\textrm{0}}$ (re-scaled over its total number of bases) and $H_{\textrm{20}}$ and an average of $H_{\textrm{40}}$ and $L_{\textrm{40}}$ (re-scaled extensions of their poly-dA loops) with their respective fits. Color code as in b. Inset shows the comparison of the theoretical re-scaled FECs for the unstacked state (dashed) and the predicted for the infinite ST-model, compared with magnetic tweezers data \cite{mcintosh2014} for a polypyrimidine and poly-dA ssDNA sequences.}
\label{fig:3}
\end{figure}

To interpret the data, we introduce a helix-coil stacking model (ST model) (Fig.~\ref{fig:4}a and Methods) where the ssDNA polymer is represented by a chain of $N$ {\it stackable} bases, that can be in the stacked (S, blue) and unstacked (U, yellow) ¡state. The elasticity of the stacked bases is given by a Worm-Like Chain (WLC) model, with the contour length $l_S$ and the persistence length $p_S$. In contrast, bases in the unstacked domains follow the TC elasticity as described in the previous Section. The Hamiltonian of the model, given by Eq.~\eqref{eq:hamiltonian_1}, depends on two energy parameters: the energy gain per stacked base, $\epsilon_{\textrm{ST}}$; and the cooperativity between neighboring bases, $\gamma_{\textrm{ST}}$. The latter is an interaction energy between adjacent bases that rewards (penalizes) bases being in the same (different) state.}¡
As schematically depicted in Fig.~\ref{fig:4}a, each stacked base contributes to the total ssDNA energy with $-\epsilon_{\textrm{ST}}$ whereas neighbouring bases in the same (different) state contribute with $-\gamma_{\textrm{ST}}$ ($+\gamma_{\textrm{ST}}$).

The model can be analytically solved for a finite chain of $N$ bases. Results for the predicted FECs for different $N$ are shown in Fig.~\ref{fig:4}b, using the parameters that best fit the experimental results (see below). For small $N$ ($N\lessapprox 10$), the unstacking transition is almost undetectable, while the shoulder in the FEC becomes visible as we approach the thermodynamic limit $N\to \infty$. The elasticities of the fully unstacked and fully stacked states are shown as black dashed and dotted lines, respectively. The model also predicts the fraction of stacked bases at a given force, $\phi_S$ (Fig.~\ref{fig:4}b, inset), which increases with $N$, saturating for $N\to \infty$. 

\begin{figure}[b]
\begin{center}
\includegraphics{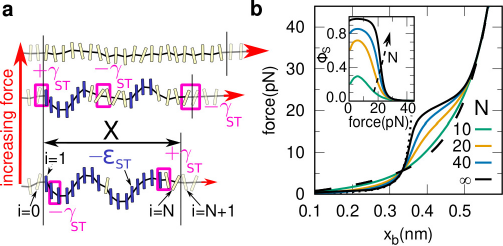}
\end{center}
\caption{{\bf Stacking (ST) model.} {\bf a.} Schematic depiction of the model for $N=20$ bases. The bases are in either a stacked (blue) or unstacked (yellow) state. The former are favored energetically by $\epsilon_{\textrm{ST}}$, while adjacent bases are energetically favored (penalized) with $\gamma_{\textrm{ST}}$ if they do (not) share state. As force increases, the longer unstacked state is energetically favored. {\bf b.} Theoretical FECs for varying lengths (color lines). Dashed and dotted lines represent the completely unstacked and unstacked elasticity, respectively. The black continuous line shows the model prediction for a domain of $N\to \infty$. The inset shows the fraction of bases in the stacked state, $\phi_S$, as a function of the force (same color code as the main panel).}
\label{fig:4}
\end{figure}
A combined fit of the ST model with 4 parameters ($l_S$, $p_S$, $\epsilon_{\textrm{ST}}$ and $\gamma_{\textrm{ST}}$) has been performed for $H_{\textrm{20}} (N=20)$ and $(HL)_{\textrm{40}} (N=40)$, giving $l_S=0.386(2)$ nm, $p_S=9.9(4)$ nm, $\epsilon_{\textrm{ST}}=0.14(1)$ kcal/mol and $\gamma_{\textrm{ST}}=0.86(2)$ kcal/mol.
The inset of Fig.~\ref{fig:3}c compares the FECs predicted by the ST-model using the obtained fitting parameters (black line) with independent experimental data from Ref.~\cite{mcintosh2014} for very long poly-dA ssDNA molecules, $N\sim 5-40$kb (black circles). The inset also compares data from Ref.~\cite{mcintosh2014} for a polypyrimidine (poly-pyr) sequence (grey circles) with the TC model prediction for the fully unstacked ssDNA at 1M NaCl (dashed black line) finding good agreement. Our values, $l_S=0.386(2)$ nm, $p_S=9.9(4)$ nm are compatible with those obtained in previous gel electrophoresis studies ($l_S\sim 0.33$ nm, $p_S\sim 7.5$ nm) \cite{Mills1999}. $l_S$ is similar to the value for poly-A RNA \cite{Seol2007prl} ($l_S=0.36$ nm), and also the step size in B-DNA (0.34nm). While $p_S$ is larger than $p_U\sim 1$ nm, it is also lower than for dsDNA ($\sim 50$ nm) at comparable salt conditions \cite{Bustamante2003,Lipfert2010}.

\begin{figure*}[b]
\begin{center}
\includegraphics{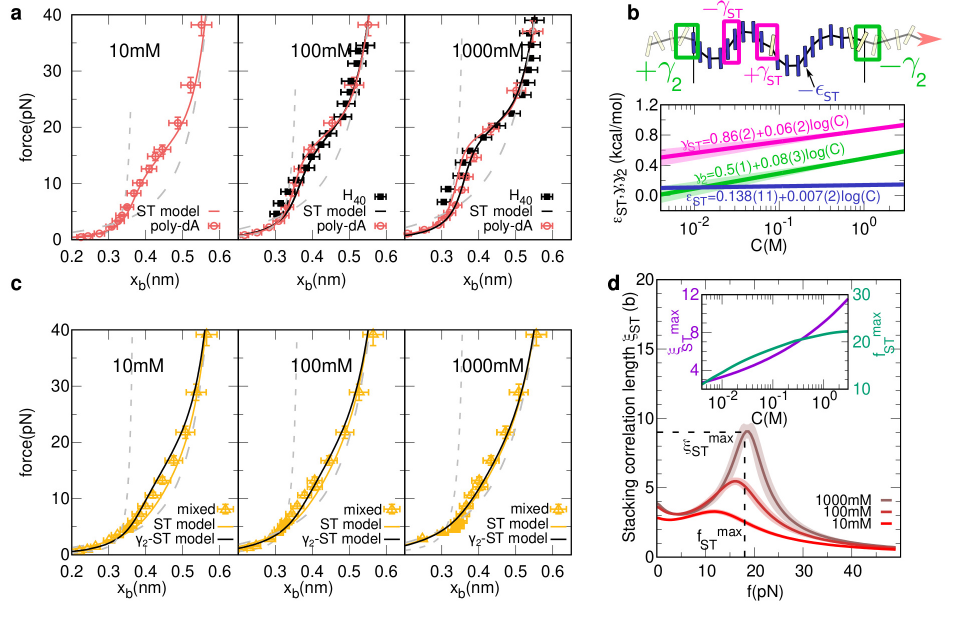}
\end{center}
\caption{{\bf Salt dependence of base stacking.} {\bf a.} FECs of the poly-dA loop for the $H_{\textrm{40}}$ (black squares) and poly-dA from Ref.~\cite{mcintosh2014} (empty red circles) for 10, 100 and 1000mM NaCl concentration. The grey dashed and dotted lines represent the elasticity of the unstacked and stacked state using the TC and WLC models, respectively. Solid curves are the fitting ones using the finite (black) and infinite (red) model (50mM and 500mM curves for $H_{\textrm{40}}$ shown in Supp. Fig. S8). {\bf b.} (top) Schematic depiction of the cooperativity between adjacent domains for the ST-$\gamma_2$ model. (bottom) Salt dependence of the stacking energy per base, $\epsilon_{\textrm{ST}}$ and interaction energies between purines ($\gamma_{\textrm{ST}}$) and the purine-pyrimidine boundary one ($\gamma_2$). {\bf c.} Results for the few kb mixed sequence of Ref.~\cite{mcintosh2014} containing the repetitive 28 b motif ({\bf AAGAG}TAT{\bf GGAAAG}T {\bf AAAAGAAA}T{\bf AAAG}) with three poly-purine regions of 9,6 and 8 b (bold letters) for 10, 100 and 1000mM NaCl (orange triangles). Solid orange and black lines correspond to the fits of the finite ST-model and the ST-$\gamma_2$ models, respectively. The grey dashed and dotted lines represent the elasticity of the unstacked and stacked state using the TC and WLC models, respectively. {\bf d} 
Correlation lengthfrom the ST model, Eq.~\eqref{eq:correlation5}, as a function of the force for 10, 100 and 1000mM NaCl concentration. Inset: Theoretical predictions for the infinite model of the maximum stacking correlation length ($\xi^{max}_{\textrm{ST}}$, magenta) and the force at which it peaks ($f^{\textrm max}_{\textrm ST}$) as a function of the salt concentration, $C$. Statistical uncertainties are shown as shadowed areas.}
\label{fig:5}
\end{figure*}

Regarding the energy parameters, the cost of a domain wall $2\gamma_{\textrm{ST}}$ is ten times the stacking energy per base $2\gamma_{\textrm{ST}} \sim 10\epsilon_{\textrm{ST}}$, highlighting the cooperativity of stacking. Interestingly, the values $\epsilon_{\textrm{ST}}=0.14(1)$ kcal/mol and $\gamma_{\textrm{ST}}=0.86(2)$ kcal/mol are similar to those for non-specific secondary structure in ssDNA, $\epsilon=0.18$ kcal/mol and $\gamma\sim 0.61$ kcal/mol at 10 mM \ce{MgCl2} \cite{viader2021prx}, suggesting that secondary structure formation is mainly driven by stacking. 

\subsection{Salt dependence of base stacking}

To further elucidate ssDNA stacking, we have investigated the effect of salt by pulling $H_{\textrm{40}}$ at different NaCl concentrations (50, 100, 500, and 1000mM, Supp. Fig. S8). Figure ~\ref{fig:5}a shows results for $H_{\textrm{40}}$ at two selected concentrations (100 and 1000mM; filled black squares in the middle and right panels). Results for 50 and 500mM are shown in Supp. Fig. S9. We compare these results to those of Ref.~\cite{mcintosh2014} on long ($N=\infty$) poly-dA sequences (empty red circles). The left panel also shows data from Ref.~\cite{mcintosh2014} for 10mM, for which we do not have data because $H_{\textrm{40}}$ does not refold in the unzipping experiments below 50mM NaCl. We have performed a simultaneous fit of the ST model combining our data for $N=40$ with data from Ref.~\cite{mcintosh2014} over the various salt conditions. 

\begin{table*}[t]

\tableparts{%
\caption{{\bf Poly-dA fitting parameters to the ST-model.} 
Values from procedure I were obtained by fitting the poly-dA data at different salt conditions and ssDNA lengths to the ST-model (single $\gamma$). Next, the mixed sequence data is fitted adding the $\gamma_2$ parameter. Values obtained from procedure II were obtained by simultaneously fitting all sequences and salt concentrations using the $\gamma_2$-ST model.}
\label{tab:table1}%
}{%
\begin{tabular*}{0.7\linewidth}{@{}cccc@{}}
\toprule
Relation (C in M units) & & Procedure I & Procedure II (kcal/mol, nm)
\\
(C in M units) & & (kcal/mol, nm) & (kcal/mol, nm)
\\
\colrule
 $\epsilon_{\textrm{ST}}=\epsilon^0_{\textrm{ST}}+m^{\epsilon}_{\textrm{ST}}\log\left(C\right)$ & & $\epsilon^{0}_{\textrm{ST}}= 0.14(1)$, $m^{\epsilon}_{\textrm{ST}}=0.004(1)$ & $\epsilon^{0}_{\textrm{ST}}= 0.13(1)$, $m^{\epsilon}_{\textrm{ST}}=0.006(2)$ \\
 $\gamma_{\textrm{ST}}=\gamma^0_{\textrm{ST}}+m^{\gamma}_{\textrm{ST}}\log\left(C\right)$ & & $\gamma^0_{\textrm{ST}}= 0.86(2)$, $m^{\gamma}_{\textrm{ST}}=0.05(1)$ & $\gamma^0_{\textrm{ST}}= 0.88(3)$, $m^{\gamma}_{\textrm{ST}}=0.05(2)$ \\
 $\gamma_{2}=\gamma^0_{2}+m_{\gamma_2}\log\left(C\right)$ & & $\gamma^0_{2}= 0.5(1)$, $m_{\gamma_2}=0.08(3)$ & $\gamma^0_{2}= 0.5(2)$, $m_{\gamma_2}=0.09(3)$ \\
 $\gamma_{2}=\gamma^0_{2}+m^{\gamma}_{\textrm{ST}}\log\left(C\right)$ & & $\gamma^0_{2}= 0.43(5)$ & \\
 $p_S=p_{S,\infty}+m_p/\sqrt{C}$ & & $p_{S,\infty}=9.9(9)$, $m_{p}=0.3(1)$ & $p_{S,\infty}=4.5(4)$, $m_{p}=0.9(5)$ \\
 $l_S=constant$ & & $l_S=0.386(2)$ & $l_S=0.40(1)$
\\
\botrule
\end{tabular*}%
}
{
}
\end{table*}

Fits have been performed by imposing a logarithmic salt dependence for the energy parameters of the model, $\epsilon_{\textrm{ST}}=\epsilon_{\textrm{ST}}^0+m_{\textrm{ST}}^{\epsilon}\log(C)$ and $\gamma_{\textrm{ST}}=\gamma_{\textrm{ST}}^0+m_{\textrm{ST}}^{\gamma}\log(C)$, with $C$ the salt concentration in M units and $\epsilon_{\textrm{ST}}^0=0.14(1)$ kcal/mol and $\gamma_{\textrm{ST}}^0=0.86(2)$ kcal/mol the reference values at 1M NaCl. These values have been imposed from the previous fits at the equivalent 10mM \ce{MgCl2} salt condition (Fig.~\ref{fig:3}c and Supp. Fig. S10). 
A logarithmic salt dependence is predicted by thermodynamic activity models of diluted ionic solutions and confirmed in studies of DNA and RNA hybridization \cite{SantaLucia1998,Huguet2010PNAS,huguet2017NAR,Bizarro2012,viader2021prx,Markham2008unafold}. A Debye-like behavior has been assumed for the persistence length of the stacked bases ($p_S=p_{\infty}+A/\sqrt{C}$ \cite{bosco2014nar}) while $l_S=0.386$ nm is taken as salt independent \cite{bosco2014nar}.

The fitting curves (Fig.~\ref{fig:5}a, continuous lines) reproduce the experimental FECs. Fig.~\ref{fig:5}b shows the salt dependence of $\epsilon_{\textrm{ST}}$ (blue) and $\gamma_{\textrm{ST}}$ (magenta), with their uncertainties (shadowed bands). We notice that $\epsilon_{\textrm{ST}}\sim 0.11-0.14$ kcal/mol ($m_{\textrm{ST}}^{\epsilon}=0.007(2)$ kcal/mol) and $p_S\sim 10-12$ nm remain almost constant, whereas $\gamma_{\textrm{ST}}$ nearly doubles from 10mM to 1M salt concentration ($m_{\textrm{ST}}^{\gamma}=0.065(17)$). Interestingly, 2$m_{\textrm{ST}}^{\gamma}=0.13(3)$ kcal/mol agrees with the salt correction energy parameter per base pair for DNA hybridization in the NN model ($0.11$ kcal/mol) \cite{SantaLucia1998,Huguet2010PNAS,huguet2017NAR}. Table~\ref{tab:table1} shows the fitting parameters obtained with the outlined procedure (procedure I, central column). 

We compare our model predictions with previous data for long ($N\to\infty$) poly-dA sequences. 
For $N=\infty$, the stacking free energy $\Delta G_0$, defined as the free energy difference between the relaxed ssDNA at zero force and the fully unstacked state is given by Eq.~\eqref{eq:free_energy}. When $\beta\gamma_{\textrm{ST}}\sim 1$ we obtain $\Delta G_0\sim \epsilon_{\textrm{ST}}+{\cal O}(e^{-4\beta\gamma_{\textrm{ST}}})\simeq 0.14(3)$ kcal/mol for all salt conditions. This value agrees with the salt independent stacking energy reported for poly-dA sequences, $\Delta G_0=0.159(13)$ kcal/mol \cite{mcintosh2014}, and is close to calorimetric and optical estimates obtained for finite $N$ sequences, $\Delta G_0=0.09-0.12$ kcal/mol \cite{Filimonov1978,Freier1981,Suurkuusk1977,Poland1966,Applequist1966,Leng1966}.

The ST model permits us to calculate the stacking correlation length $\xi_{\textrm{ST}}$ versus force $f$, Eq.~\eqref{eq:correlation5} (Fig.~\ref{fig:5}d at 0.01, 0.1, 1M \ce{NaCl}). The stacking correlation length sets the minimum nucleation size that triggers stacked domain growth. It shows a maximum $\xi^{max}_{\textrm{ST}}$ at $f^{\textrm max}_{\textrm ST}$, dropping to zero at high forces. $\xi_{\textrm{ST}}^{max}$ is salt dependent varying from 4b at 10mM to 10b at 1M (Fig.~\ref{fig:5}d, inset). The maximum $\xi_{\textrm{ST}}^{max}$ is a consequence of the first order character of the stacking-unstacking transition. The larger extensional fluctuations in the pulling direction are due to the breathing of the planes of the bases leading to the shoulder observed in the FECs (e.g., Fig.~\ref{fig:3}c). One can prove that the maximum in the correlation length $\xi_{\textrm{ST}}^{max}$ occurs at a force $f^{\textrm max}_{\textrm ST}$ that depends on $\epsilon_{\textrm ST}$ and the elasticity of the stacked and unstacked states, $\epsilon_{\textrm ST}=\int_{0}^{f^{max}} (x_U(f')-x_S(f')) df’$. 
In contrast, the value $\xi_{\textrm{ST}}^{max}$ only depends on the cooperativity parameter $\gamma_{\textrm ST}$, $\xi_{\textrm{ST}}^{max}=-1/(\log(\tanh(\beta\gamma_{\textrm ST}))$. Both $\xi_{\textrm{ST}}^{max}$ and $f^{\textrm max}_{\textrm ST}$ increase with salt concentration. These results are general predictions of helix-coil models and are derived in Supp. Sec. 5.1. Interestingly, our predicted value $\xi_{\textrm{ST}}(f=0)=4$b matches the minimal nucleation length reported in calorimetry experiments \cite{ramprakash2008biopol}. The value of $f^{max}=14-20$ pN also matches the shoulder observed in the FECs; the larger the $\xi_{\textrm{ST}}^{max}$, the more prominent the shoulder is (Figs.~\ref{fig:3}c and \ref{fig:5}a). $f^{\textrm max}_{\textrm ST}$ also agrees with predictions based on electrostatic tension models \cite{mcintosh2014,Netz2001,Manning2006} (Supp. Sec. 6 and Supp. Fig. S11).

Stacking also occurs between purines and pyrimidines \cite{ramprakash2008biopol}. To investigate purine-pyrimidine stacking, we have extended the ST model (Eq.\ref{eq:hamiltonian_1}) by considering purine-like ({\it stackable}) and pyrimidine-like ({\it non-stackable}) domains and introducing a cooperativity parameter $\gamma_2$ at the purine-pyrimidine boundaries (Fig.~\ref{fig:5}b, green boxes in top schematics). The FECs predicted by the $\gamma_2$-ST model (Supp. Sec. 5C) have been fitted to pulling data from Ref.~\cite{mcintosh2014} for mixed sequences containing tracks of 6, 8, 9 consecutive purines (Fig.~\ref{fig:5}c, triangles). We have assumed the logarithmic salt dependence, $\gamma_2=\gamma_2^{0}+m_{\gamma_2}\log{C}$ with the previously determined parameters $\epsilon_{\textrm{ST}}$, $\gamma_{\textrm{ST}}$, $p_S$, $l_S$, $p_U$, $l_U$. We obtain $\gamma_2=0.5(1)+0.08(3)\log{C}$ (kcal/mol) which is lower than $\gamma_{\textrm{ST}}$ (kcal/mol) (green and pink lines in Fig.~\ref{fig:5}b) indicating lower stacking cooperativity between purines and pyrimidines. Interestingly, the salt correction parameters for $\gamma_2$ and $\gamma_{\textrm{ST}}$ are close (0.065 versus 0.08), indicating similar ion activity effects for stacking. In fact, by imposing the salt correction parameter $m_{\textrm{ST}}^{\gamma}$ to the fit of $\gamma_2$ we get compatible results, $\gamma_2^0=0.43(5)$ kcal/mol. Table~\ref{tab:table1} shows the parameters obtained by a simultaneous fit of all data to the $\gamma_2$-ST model (procedure II, right column) and agrees with the results of the previous analysis (procedure I, middle column). 

\subsection{Base-stacking in other purine sequences}
\begin{figure}[t]
\begin{center}
\includegraphics{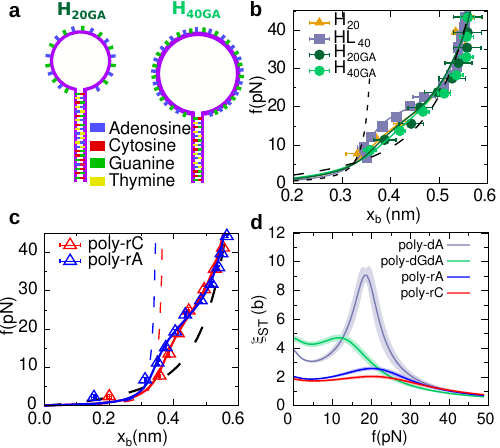}
\end{center}
\caption{{\bf Stacking in ssRNA compared to ssDNA.} {\bf a.} Schematic depictions of the two hairpins with poly-dGdA motifs in the loop. {\bf b.} Re-scaled FECs per base for the dGdA loops of $H_{\textrm{20GA}}$ (dark green), $H_{\textrm{40GA}}$ (light green). As a comparison, data for the 20b and 40b dA loops (orange and gray) is shown. Dashed lines represent the elasticity of the stacked and unstacked conformations. Continuous lines are fits of the ST model to the experimental data. {\bf c.} Re-scaled FECs per base for poly-rA (blue) and poly-rC (red) sequences from Reference \cite{Seol2007prl}. Continuous lines are fits of the ST model to the experimental data. In contrast, dashed lines represent the elasticity of the stacked (blue for poly-rA and red for poly-rC) and unstacked (black) conformations. {\bf d} Correlation length from the ST model, Eq.~\eqref{eq:correlation5}, as a function of the force for poly-dA, poly-dGdA, poly-rA and polyrC. Statistical uncertainties are shown as shadowed areas. 
Error bars are the statistical errors from averaging different molecules (Supp. Table S6).}
\label{fig:6}
\end{figure}

To further investigate stacking in ssDNA, we have considered poly-dGdA (alternating dA and dG bases) and poly-dG sequences of different lengths embedded in the same hairpin stems shown in Figure~\ref{fig:1}. 
Hairpins $H_{\textrm 20GA}$ and $H_{\textrm 20G}$ have a stem of 20bp and a loop of 20 bases, whereas hairpins $H_{\textrm 40GA}$ and $H_{\textrm 40G}$ have a stem of 15bp and a loop of 40 bases, so the total number of bases are 60 and 70, respectively. Hairpin sequences are shown in Figure~\ref{fig:6}a, Supp. Fig. S12 and Supp. Table S1.
We have pulled the four new constructs (Supp. Figs. S12, S13 and S14). Results on poly-dGdA sequences ($H_{\textrm{20GA}}$ and $H_{\textrm{40GA}}$) are shown in Fig.~\ref{fig:6}b as green circles and are compared to poly-dA results. A small deviation from the ideal elastic behavior (dashed line) and a a modest shoulder suggest that stacking is weaker for poly-dGdA than for poly-dA. Fits of the ST model are shown as continuous lines. The weaker stacking in poly-dGdA is reflected in the fitting parameters, which give $\epsilon_{\textrm ST}=0.02(1)$ kcal/mol and $\gamma_{\textrm ST}=0.67(2)$ kcal/mol, compared to $\epsilon_{\textrm{ST}}=0.14(1)$ kcal/mol and $\gamma_{\textrm{ST}}=0.86(2)$ kcal/mol for poly-d, leading to a lower free energy of $\Delta G_0^{ST}=0.07(3)$ kcal/mol (as compared to $\Delta G_0^{ST}=0.14(3)$ for poly-dA). 
Results for the force-dependent correlation length $\xi_{\textrm ST}$ are shown in Fig.~\ref{fig:6}d. Compared to poly-dA, the correlation length for poly-dGdA shows a less pronounced maximum at a lower force, $12$ pN versus 18pN, which is a consequence of the lower stacking cooperativity ($\gamma_{\textrm ST}$) and stability ($\epsilon_{\textrm ST}$) in poly-dGdA. The lower $\xi^{max}_{\textrm{ST}}$ aligns with the negligible finite-size effects observed in the FECs of $H_{\textrm{20GA}}$ and $H_{\textrm{40GA}}$, dark and light green color circles in Fig.~\ref{fig:6}b.

Finally, pulling experiments on the poly-dG constructs lead to non-reproducible FECs that we interpret as due to the formation of compact structures, such as G-quadruplexes, that unfold at forces higher than 40pN. The remarkable kinetic stability of such structures precludes stacking measurements of poly-dG sequences using our method (see Supp. Figure S14.).

\subsection{Stacking of ssRNA}

Our study of poly-dA naturally extends to poly-rA sequences, relevant for the tailing of mRNA during the maturation process \cite{Passmore2022}. Poly-rA tails contain hundreds of rA bases that confer a high rigidity to the backbone potentially influencing mRNA translation and gene expression \cite{Tang2020}. Previous force-spectroscopy measurements in few kilobases homopolymeric ssRNA molecules revealed a stacking-unstacking transition with the characteristic FEC shoulder \cite{Seol2007prl}. While no stacking was observed for poly-U, stacking was observed for poly-rA and poly-rC. The poly-U FEC is well described by the ssDNA unstacked elasticity of the Thick-Chain model, as shown in Supp. Fig. S15. We have analyzed the data of poly-rA and poly-rC from Ref.~\cite{Seol2007prl} with our ST-model ($N=\infty$) successfully reproducing the data, Fig.~\ref{fig:6}c. We find that the stacking energy parameter $\epsilon_{\textrm ST}$ is larger for poly-rA than for poly-dA (0.18kcal/mol vs. 0.14kcal/mol), in agreement with the higher value of the force where the shoulder occurs in the FEC is observed for poly-rA. On the other hand, the cooperativity parameter $\gamma_{\textrm ST}$ is lower for poly-rA than for poly-dA (0.5 kcal/mol vs. 0.8 kcal/mol), which results in a shorter stacking correlation length, as shown in Fig.~\ref{fig:6}d. Besides, the maximum correlation length is observed at larger forces ($\sim$20-25 pN, Fig. ~\ref{fig:6}d), in agreement with the larger stacking-unstacking transition force. In contrast, the free energy of stacking per base of poly-rA at zero force, Eq.~\ref{eq:free_energy}, is 1.6 times larger than that of poly-dA. For poly-rC, we obtain a lower $\epsilon_{\textrm ST}=0.13$ kcal/mol than that for poly-rA but a similar $\gamma_{\textrm ST}=0.4$ kcal/mol, leading to a lower correlation length (Fig. \ref{fig:6}d) and stacking free energy ($\Delta G_0^{ST}=0.20$ kcal/mol for poly-rC vs $\Delta G_0^{ST}=0.25$ kcal/mol for poly-rA).

\section{CONCLUSIONS}

Base pairing and stacking are recognized as the main driving forces in nucleic acids folding. While Watson-Crick base pairing is key to modeling specific secondary structures, stacking is less specific and non-local, tending to pile up bases along molecular chains. 
The cumulative effect of several stacked bases does lead to cooperative and collective effects. 
Despite their importance, stacking energies in ssNA are poorly known due to their low values, about $\sim0.1$ kcal/mol per base. Here, we have applied the blocking-splint oligo method to accurately measure the mechanical response of poly-dA tracks of 20-40 bases in a broad range of forces and salt conditions using pulling experiments. 
A helix-coil model for stacking reproduces the experimentally measured force-extension curves, showing finite-size effects. Such effects are due to the finite stacking correlation length $\xi_{\textrm{ST}}\sim 5-10$b, on the scale of 20-40b of the poly-dA loops studied in the paper. We find that cooperativity increases with salt concentration, doubling from $\gamma_{\textrm{ST}}\sim 0.5$ to $\sim 0.9$ kcal/mol from $10$mM to $1$M NaCl. Cooperativity is ten times larger than the energy parameter $\epsilon_{\textrm{ST}}\sim 0.1$ kcal/mol and the stacking free energy per base (Eq.~\eqref{eq:free_energy}) $\Delta G_0^{ST}\simeq 0.14(3)$ kcal/mol, which are nearly salt independent. 
These results suggest that cooperativity is salt-dependent, despite that stacking stability remains salt-independent, in agreement with previous results \cite{mcintosh2014}. Consequently, the shoulder of the FECs of poly-dA tracks becomes more prominent with salt, while the area between the FECs and the unstacked elastic response remains constant (Supp. Fig. S16). Moreover, the salt correction parameter for DNA hybridization in the NN model ($\sim0.11$ kcal/mol) \cite{SantaLucia1998,Huguet2010PNAS,huguet2017NAR} equals twice the salt correction parameter for $\gamma_{\textrm ST}$, $2m_{\textrm{ST}}^{\gamma}=0.13(3)$ kcal/mol, suggesting that double helix stability is mainly due to stacking, in agreement with other studies \cite{Protozanova2004,Yakovchuk2006}.
Remarkably, the measured elasticities in 10mM \ce{MgCl2} and 1M \ce{NaCl} are indistinguishable, indicating that the 1:100 salt rule-of-thumb holds for the stacking-unstacking transition. Therefore, the stacking cooperativity, stability and correlation length can be extrapolated to physiological conditions \cite{Melkikh2008,Szatmari2020}, of equivalent ionic strength $\sim150-250$mM \ce{NaCl}: $\gamma_{\textrm ST}=0.7$ kcal/mol, $\epsilon_{\textrm ST}=0.14$ kcal/mol, $\xi_{\textrm ST}=7-8$ b. 

We have also studied other purine sequences, such as poly-dGdA tracks of weaker stacking showing a lower $\epsilon_{\textrm ST}$, $\gamma_{\textrm ST}$ and $\Delta G_0^{ST}$, leading to shorter $\xi_{\textrm ST}$ at the stacking-unstacking transition force and negligible finite-size effects. Overall, the poly-purine sequences studied present a strong stacking cooperativity with a correlation length of around 4b at zero force.

Besides, the formation of stable G-quadruplexes-like structures precludes stacking measurements in poly-dG sequences using our approach.
Finally, we have applied the ST model to homopolymeric ssDNA and ssRNA sequences studied in previous works \cite{Seol2007prl,mcintosh2014}, finding that stacking energies are larger for ssRNA. In contrast, the cooperativity and stacking correlation length are lower.

Stacking cooperativity is crucial in nucleic acids. In duplex DNA, stacking is responsible for the allosteric effects \cite{Sangjin2013} that propagate long-range interactions in ligand binding, important for regulating gene expression \cite{Rosenblum2021}. Mechanical models with cooperativity find collective binding affinities of periodicity equal to the helical pitch \cite{Carlon2023}. Cooperativity effects in the form of stacked base triplets have also been observed in overstretched DNA that might be related to the triplets of the genetic code \cite{Bosaeus2017}. Besides, RNA folds cooperatively into large tertiary structures stabilized by water bridges between phosphates and bases and additional inter-strand stacks.
Previous studies of poly-rA RNA molecules \cite{Leng1966,Rich1961,Safaee2013} showed that they form double-stranded helices stabilized by stacking, despite being unable to form Watson-Crick base pairs. 
We generally expect stacking interactions to be more important than hydrogen bonding for cooperativity effects in duplex and ssNA since stacking is the only force that naturally propagates along the phosphate backbone.
Our results ($\gamma_{\textrm ST}>>\epsilon_{\textrm ST}$) support the relevance of stacking cooperativity in ssNA in promoting different kinds of structures.
The persistence and stacking correlation lengths in ssNA are central parameters for understanding hybridization and assembly of ssNA sequences, a key process for synthetic devices such as DNA origami \cite{Castro2011}, DNA nanoswitches \cite{Green2019} and synthetic molecular machines \cite{Balzani2000}.

Moreover, the stacking properties of ssDNA regulate the binding affinity of different single-stranded binding (SSB) proteins involved in replication and recombination \cite{Kim1995}. 
The distinct elastic properties of homopolymeric sequences imply different affinities of regulatory proteins that can be characterized using high-throughput techniques such as FRET platforms \cite{Hartmann2023} and electrostatic traps \cite{Chu2024}. Moreover, the studies of homopolymeric sequences could be used as labels or targets for recognizing specific sequences, such as in immune detection \cite{Assi2018}. Overall, the distinct behaviors of homopolymeric single-stranded sequences might have been important in codifying specific functionalities in some stages of evolution.

Remarkably, in recent work, we have shown that RNA, but not DNA, exhibits novel properties at cold temperatures below 20ºC attributed to the  to base-pairing interactions \cite{Rissone2024}.It would be interesting to measure the temperature-dependent stacking in ssRNA and ssDNA to search for differences in the ribose-deoxyribose replacement. Measurements varying the temperature would  also allow to determine the entropy, enthalpy, and $\Delta C_p$ of stacking.

\section{FUNDING AND ACKNOWLEDGEMENTS}

This work was supported by the Fondazione Cariparo Visiting Programme 2018 (TIRES) and the project {\it Study of the elastic properties of a DNA biosensor by using optical tweezers} (BIRD207923-2020); the Spanish Research Council Grant PID2019-111148GB-100; Grant CNS2022-135910 funded by MICIU/AEI/10.13039/501100011033 and by NextGenerationEU/PRTR; the ICREA Academia 2018 and the Spanish Ramon y Cajal program of MICINN. Authors thank C. Micheletti and O. Saleh for their insightful comments about the article.

\subsubsection{Conflict of interest statement.} None declared.
\newpage

\bibliographystyle{unsrt}
\bibliography{bibliography.bib}

\begin{thebibliography}{10}

\bibitem{saenger1984}
W.~Saenger.
\newblock {\em Principles of Nucleic Acid Structure}.
\newblock Springer Science \& Business Media, 2013.

\bibitem{Holbrook1999}
Jill~A. Holbrook, Michael~W. Capp, Ruth~M. Saecker, and M.~Thomas Record.
\newblock Enthalpy and heat capacity changes for formation of an oligomeric
  {DNA} duplex: Interpretation in terms of coupled processes of formation and
  association of single-stranded helices.
\newblock {\em Biochemistry}, 38:8409--8422, 6 1999.

\bibitem{Whitley2017}
Kevin~D. Whitley, Matthew~J. Comstock, and Yann~R. Chemla.
\newblock Elasticity of the transition state for oligonucleotide hybridization.
\newblock {\em Nucleic Acids Research}, 45:547--555, 1 2017.

\bibitem{Kim1995}
Changsoo Kim and Marc~S. Wold.
\newblock Recombinant human replication protein a binds to polynucleotides with
  low cooperativity.
\newblock {\em Biochemistry}, 34(6):2058--2064, Feb 1995.

\bibitem{Goldman2015}
Daniel~H. Goldman, Christian~M. Kaiser, Anthony Milin, Maurizio Righini,
  Ignacio Tinoco, and Carlos Bustamante.
\newblock Mechanical force releases nascent chain–mediated ribosome arrest in
  vitro and in vivo.
\newblock {\em Science}, 348(6233):457--460, 2015.

\bibitem{bustamante_yan_2022}
Carlos Bustamante and Shannon Yan.
\newblock The development of single molecule force spectroscopy: from polymer
  biophysics to molecular machines.
\newblock {\em Quarterly Reviews of Biophysics}, 55:e9, 2022.

\bibitem{Protozanova2004}
Ekaterina Protozanova, Peter Yakovchuk, and Maxim~D. Frank-Kamenetskii.
\newblock Stacked-unstacked equilibrium at the nick site of {DNA}.
\newblock {\em journal of Molecular Biology}, 342:775--785, 9 2004.

\bibitem{Yakovchuk2006}
Peter Yakovchuk, Ekaterina Protozanova, and Maxim~D. Frank-Kamenetskii.
\newblock Base-stacking and base-pairing contributions into thermal stability
  of the {DNA} double helix.
\newblock {\em Nucleic Acids Research}, 34:564--574, 2 2006.

\bibitem{viader2021prx}
X.~Viader-Godoy, C.~R. Pulido, B.~Ibarra, M.~Manosas, and F.~Ritort.
\newblock Cooperativity-dependent folding of single-stranded {DNA}.
\newblock {\em Phys. Rev. X}, 11:031037, Aug 2021.

\bibitem{Chakraborty2009}
Saikat Chakraborty, Suruchi Sharma, Prabal~K Maiti, and Yamuna Krishnan.
\newblock The poly d{A} helix: a new structural motif for high performance
  {DNA-based} molecular switches.
\newblock {\em Nucleic Acids Res}, 37(9):2810--2817, March 2009.

\bibitem{Rich1961}
Alexander Rich, David~R. Davies, F.H.C. Crick, and J.D. Watson.
\newblock The molecular structure of polyadenylic acid.
\newblock {\em Journal of Molecular Biology}, 3(1):71--IN19, 1961.

\bibitem{Gehring1993}
Kalle Gehring, Jean-Louis Leroy, and Maurice Gu{\'e}ron.
\newblock A tetrameric {DNA} structure with protonated cytosine-cytosine base
  pairs.
\newblock {\em Nature}, 363(6429):561--565, Jun 1993.

\bibitem{Assi2018}
Hala Abou~Assi, Miguel Garav{\'i}s, Carlos Gonz{\'a}lez, and Masad~J. Damha.
\newblock i-motif {DNA}: structural features and significance to cell biology.
\newblock {\em Nucleic Acids Research}, 46(16):8038--8056, Sep 2018.

\bibitem{Robinson2021}
Jenna Robinson, Federica Raguseo, Sabrina~Pia Nuccio, Denise Liano, and Marco
  Di~Antonio.
\newblock {{DNA} G-quadruplex structures: more than simple roadblocks to
  transcription?}
\newblock {\em Nucleic Acids Research}, 49(15):8419--8431, 07 2021.

\bibitem{Castro2011}
Carlos~Ernesto Castro, Fabian Kilchherr, Do-Nyun Kim, Enrique~Lin Shiao, Tobias
  Wauer, Philipp Wortmann, Mark Bathe, and Hendrik Dietz.
\newblock A primer to scaffolded {DNA} origami.
\newblock {\em Nature Methods}, 8(3):221--229, Mar 2011.

\bibitem{Green2019}
Leopold~N.F Green, Hari K.~K. Subramanian, Vahid Mardanlou, Jongmin Kim,
  Rizal~F. Hariadi, and Elisa Franco.
\newblock Autonomous dynamic control of {DNA} nanostructure self-assembly.
\newblock {\em Nature Chemistry}, 11(6):510--520, Jun 2019.

\bibitem{Balzani2000}
Vincenzo Balzani, Alberto Credi, Françisco~M. Raymo, and J.~Fraser Stoddart.
\newblock Artificial molecular machines.
\newblock {\em Angewandte Chemie International Edition}, 39(19):3348--3391, Oct
  2000.

\bibitem{Kilchherr2016}
Fabian Kilchherr, Christian Wachauf, Benjamin Pelz, Matthias Rief, Martin
  Zacharias, and Hendrik Dietz.
\newblock Single-molecule dissection of stacking forces in {DNA}.
\newblock {\em Science}, 353(6304):aaf5508, 2016.

\bibitem{SantaLucia1998}
John SantaLucia.
\newblock A unified view of polymer, dumbbell, and oligonucleotide {DNA}
  nearest-neighbor thermodynamics.
\newblock {\em Proceedings of the National Academy of Sciences},
  95(4):1460--1465, 1998.

\bibitem{Huguet2010PNAS}
Josep~M. Huguet, Cristiano~V. Bizarro, N{\'u}ria Forns, Steven~B. Smith, Carlos
  Bustamante, and Felix Ritort.
\newblock Single-molecule derivation of salt dependent base-pair free energies
  in {DNA}.
\newblock {\em Proceedings of the National Academy of Sciences},
  107(35):15431--15436, 2010.

\bibitem{huguet2017NAR}
Josep~Maria Huguet, Marco Ribezzi-Crivellari, Cristiano~Valim Bizarro, and
  Felix Ritort.
\newblock Derivation of nearest-neighbor {DNA} parameters in magnesium from
  single molecule experiments.
\newblock {\em Nucleic Acids Research}, 45(22):12921--12931, 2017.

\bibitem{Gray1970}
Donald~M. Gray and Ignacio Tinoco~Jr.
\newblock A new approach to the study of sequence-dependent properties of
  polynucleotides.
\newblock {\em Biopolymers}, 9(2):223--244, 1970.

\bibitem{Licinio2007}
Pedro Licinio and João Carlos~O. Guerra.
\newblock Irreducible representation for nucleotide sequence physical
  properties and self-consistency of nearest-neighbor dimer sets.
\newblock {\em Biophysical journal}, 92(6):2000--2006, 2007.

\bibitem{Bosaeus2012}
Niklas Bosaeus, Afaf~H. El-Sagheer, Tom Brown, Steven~B. Smith, Björn
  Åkerman, Carlos Bustamante, and Bengt Nordén.
\newblock Tension induces a base-paired overstretched dna conformation.
\newblock {\em Proceedings of the National Academy of Sciences},
  109(38):15179--15184, 2012.

\bibitem{Feng2019}
Bobo Feng, Robert~P. Sosa, Anna K.~F. Mårtensson, Kai Jiang, Alex Tong,
  Kevin~D. Dorfman, Masayuki Takahashi, Per Lincoln, Carlos~J. Bustamante,
  Fredrik Westerlund, and Bengt Nordén.
\newblock Hydrophobic catalysis and a potential biological role of {DNA}
  unstacking induced by environment effects.
\newblock {\em Proceedings of the National Academy of Sciences},
  116(35):17169--17174, 2019.

\bibitem{Punnoose2023}
Jibin Abraham~Punnoose, Kevin~J. Thomas, Arun~Richard Chandrasekaran, Javier
  Vilcapoma, Andrew Hayden, Kacey Kilpatrick, Sweta Vangaveti, Alan Chen,
  Thomas Banco, and Ken Halvorsen.
\newblock High-throughput single-molecule quantification of individual base
  stacking energies in nucleic acids.
\newblock {\em Nature Communications}, 14(1):631, Feb 2023.

\bibitem{ramprakash2008biopol}
Jayanthi Ramprakash, Brian Lang, and Frederick~P. Schwarz.
\newblock Thermodynamics of single strand {DNA} base stacking.
\newblock {\em Biopolymers}, 89(11):969--979, 2008.

\bibitem{Isaksson2004}
J.~Isaksson, S.~Acharya, J.~Barman, P.~Cheruku, and J.~Chattopadhyaya.
\newblock Single-stranded adenine-rich {DNA} and {RNA} retain structural
  characteristics of their respective double-stranded conformations and show
  directional differences in stacking pattern.
\newblock {\em Biochemistry}, 43(51):15996--16010, Dec 2004.

\bibitem{plumridge2017}
Alex Plumridge, Steve~P. Meisburger, Kurt Andresen, and Lois Pollack.
\newblock The impact of base stacking on the conformations and electrostatics
  of single-stranded {DNA}.
\newblock {\em Nucleic Acids Res}, 45(7):3932--3943, Apr 2017.
\newblock 28334825[pmid].

\bibitem{Aalberts2003}
Daniel~P. Aalberts, John~M. Parman, and Noel~L. Goddard.
\newblock Single-strand stacking free energy from {DNA} beacon kinetics.
\newblock {\em Biophysical journal}, 84(5):3212--3217, 2003.

\bibitem{Ke2007}
Changhong Ke, Michael Humeniuk, Hanna S-Gracz, and Piotr~E. Marszalek.
\newblock Direct measurements of base stacking interactions in {DNA} by
  single-molecule atomic-force spectroscopy.
\newblock {\em Phys. Rev. Lett.}, 99:018302, Jul 2007.

\bibitem{mcintosh2014}
DustinB. McIntosh, Gina Duggan, Quentin Gouil, and OmarA. Saleh.
\newblock Sequence-dependent elasticity and electrostatics of single-stranded
  {DNA}: Signatures of base-stacking.
\newblock {\em Biophysical journal}, 106(3):659 -- 666, 2014.

\bibitem{Seol2007prl}
Yeonee Seol, Gary~M. Skinner, Koen Visscher, A{RNA}ud Buhot, and Avraham
  Halperin.
\newblock Stretching of homopolymeric {RNA} reveals single-stranded helices and
  base-stacking.
\newblock {\em Phys. Rev. Lett.}, 98:158103, Apr 2007.

\bibitem{Smith2003}
Steven~B. Smith, Yujia Cui, and Carlos Bustamante.
\newblock {\em Optical-trap force transducer that operates by direct
  measurement of light momentum}, volume 361 of {\em Biophotonics, Part B},
  pages 134--162.
\newblock Academic Press, Jan 2003.

\bibitem{gieseler2020optical}
Jan Gieseler, Juan~Ruben Gomez-Solano, Alessandro Magazz\`{u}, Isaac~P\'{e}rez
  Castillo, Laura~P\'{e}rez Garc\'{i}a, Marta Gironella-Torrent, Xavier
  Viader-Godoy, Felix Ritort, Giuseppe Pesce, Alejandro~V. Arzola, Karen
  Volke-Sep\'{u}lveda, and Giovanni Volpe.
\newblock Optical tweezers --- from calibration to applications: a tutorial,
  Mar 2021.

\bibitem{Zaltron2020EPJP}
Annamaria Zaltron, Michele Merano, Giampaolo Mistura, Cinzia Sada, and Flavio
  Seno.
\newblock Optical tweezers in single-molecule experiments.
\newblock {\em European Physical journal Plus}, 135(11), 2020.

\bibitem{Landuzzi2020}
F.~Landuzzi, X.~Viader-Godoy, F.~Cleri, I.~Pastor, and F.~Ritort.
\newblock Detection of single {DNA} mismatches by force spectroscopy in short
  {DNA} hairpins.
\newblock {\em The journal of Chemical Physics}, 152:074204, 2 2020.

\bibitem{Zhang2009science}
Xiaohui Zhang, Kenneth Halvorsen, Cheng-Zhong Zhang, Wesley~P. Wong, and
  Timothy~A. Springer.
\newblock Mechanoenzymatic cleavage of the ultralarge vascular protein von
  willebrand factor.
\newblock {\em Science}, 324(5932):1330--1334, 2009.

\bibitem{viader2021ijms}
Xavier Viader-Godoy, Maria Manosas, and Felix Ritort.
\newblock Sugar-pucker force-induced transition in single-stranded {DNA}.
\newblock {\em International journal of Molecular Sciences}, 22(9), 2021.

\bibitem{alemany2014biopol}
Anna Alemany and Felix Ritort.
\newblock Determination of the elastic properties of short {ss{DNA}} molecules
  by mechanically folding and unfolding {DNA} hairpins.
\newblock {\em Biopolymers}, 101(12):1193--1199, 2014.

\bibitem{marko1995stretching}
John~F Marko and Eric~D Siggia.
\newblock Stretching {DNA}.
\newblock {\em Macromolecules}, 28(26):8759--8770, 1995.

\bibitem{Manosas2017}
Maria Manosas, Joan Camunas-Soler, Vincent Croquette, and Felix Ritort.
\newblock Single molecule high-throughput footprinting of small and large {DNA}
  ligands.
\newblock {\em Nature Communications}, 8(1):304, Aug 2017.

\bibitem{bosco2014nar}
Alessandro Bosco, Joan Camunas-Soler, and Felix Ritort.
\newblock Elastic properties and secondary structure formation of
  single-stranded {DNA} at monovalent and divalent salt conditions.
\newblock {\em Nucleic acids research}, 42(3):2064--2074, 2014.

\bibitem{Rissone2022}
Paolo Rissone, Cristiano~V. Bizarro, and Felix Ritort.
\newblock Stem–loop formation drives {RNA} folding in mechanical unzipping
  experiments.
\newblock {\em Proceedings of the National Academy of Sciences},
  119(3):e2025575119, 2022.

\bibitem{toan2005biophys}
Ngo~Minh Toan, Davide Marenduzzo, and Cristian Micheletti.
\newblock Inferring the diameter of a biopolymer from its stretching response.
\newblock {\em Biophysical journal}, 89(1):80--86, 2005.

\bibitem{Toan2006}
Ngo~Minh Toan and Cristian Micheletti.
\newblock Inferring the effective thickness of polyelectrolytes from stretching
  measurements at various ionic strengths: applications to {DNA} and {RNA}.
\newblock {\em journal of Physics: Condensed Matter}, 18(14):S269, mar 2006.

\bibitem{camunas2016annurev}
Joan Camunas-Soler, Marco Ribezzi-Crivellari, and Felix Ritort.
\newblock Elastic properties of nucleic acids by single-molecule force
  spectroscopy.
\newblock {\em Annual review of biophysics}, 45(1):65--84, 2016.

\bibitem{Bustamante2021}
Carlos~J. Bustamante, Yann~R. Chemla, Shixin Liu, and Michelle~D. Wang.
\newblock Optical tweezers in single-molecule biophysics.
\newblock {\em Nature Reviews Methods Primers}, 1(1):25, Mar 2021.

\bibitem{Mills1999}
Janine~B Mills, Elsi Vacano, and Paul~J Hagerman.
\newblock Flexibility of single-stranded {DNA}: use of gapped duplex helices to
  determine the persistence lengths of poly(d{T}) and poly(d{A}).
\newblock {\em journal of Molecular Biology}, 285(1):245--257, 1999.

\bibitem{Bustamante2003}
Carlos Bustamante, Zev Bryant, and Steven~B. Smith.
\newblock Ten years of tension: single-molecule {DNA} mechanics.
\newblock {\em Nature}, 421(6921):423--427, Jan 2003.

\bibitem{Lipfert2010}
Jan Lipfert, Sven Klijnhout, and Nynke~H. Dekker.
\newblock {Torsional sensing of small-molecule binding using magnetic
  tweezers}.
\newblock {\em Nucleic Acids Research}, 38(20):7122--7132, 07 2010.

\bibitem{Bizarro2012}
Cristiano~V Bizarro, Anna Alemany, and Felix Ritort.
\newblock Non-specific binding of {Na+} and {Mg2+} to {RNA} determined by force
  spectroscopy methods.
\newblock {\em Nucleic acids research}, 40(14):6922--6935, 2012.

\bibitem{Markham2008unafold}
Nicholas~R Markham and Michael Zuker.
\newblock {UNAFold}: software for nucleic acid folding and hybridization.
\newblock {\em Methods Mol Biol}, 453:3--31, 2008.

\bibitem{Filimonov1978}
V~V Filimonov and P~L Privalov.
\newblock Thermodynamics of base interaction in (a)n and (a.u)n.
\newblock {\em journal of molecular biology}, 122:465--70, 7 1978.

\bibitem{Freier1981}
S~M Freier, K~O Hill, T~G Dewey, L~A Marky, K~J Breslauer, and D~H Turner.
\newblock Solvent effects on the kinetics and thermodynamics of stacking in
  poly(cytidylic acid).
\newblock {\em Biochemistry}, 20:1419--26, 3 1981.

\bibitem{Suurkuusk1977}
J~Suurkuusk, J~Alvarez, E~Freire, and R~Biltonen.
\newblock Calorimetric determination of the heat capacity changes associated
  with the conformational transitions of polyriboadenylic acid and
  polyribouridylic acid.
\newblock {\em Biopolymers}, 16:2641--52, 12 1977.

\bibitem{Poland1966}
D~Poland, J~N Vou{RNA}kis, and H~A Scheraga.
\newblock Cooperative interactions in single-strand oligomers of adenylic acid.
\newblock {\em Biopolymers}, 4:223--35, 1966.

\bibitem{Applequist1966}
J~Applequist and V~Damie.
\newblock Thermodynamics of the one-stranded helix-coil equilibrium in
  polyadenylic acid.
\newblock {\em journal of the American Chemical Society}, 88:3895--900, 9 1966.

\bibitem{Leng1966}
M~Leng and G~Felsenfeld.
\newblock A study of polyadenylic acid at neutral ph.
\newblock {\em journal of molecular biology}, 15:455--66, 2 1966.

\bibitem{Netz2001}
Roland~R. Netz.
\newblock Strongly stretched semiflexible extensible polyelectrolytes and
  {DNA}.
\newblock {\em Macromolecules}, 34(21):7522--7529, Oct 2001.

\bibitem{Manning2006}
Gerald~S. Manning.
\newblock The persistence length of {DNA} is reached from the persistence
  length of its null isomer through an inte{RNA}l electrostatic stretching
  force.
\newblock {\em Biophysical journal}, 91(10):3607--3616, 2006.

\bibitem{Passmore2022}
Lori~A. Passmore and Jeff Coller.
\newblock Roles of mrna poly(a) tails in regulation of eukaryotic gene
  expression.
\newblock {\em Nature Reviews Molecular Cell Biology}, 23(2):93--106, Feb 2022.

\bibitem{Tang2020}
Terence T~L Tang and Lori~A Passmore.
\newblock Recognition of {Poly(A}) {RNA} through its intrinsic helical
  structure.
\newblock {\em Cold Spring Harb Symp Quant Biol}, 84:21--30, April 2020.

\bibitem{Melkikh2008}
A.V. Melkikh and M.I. Sutormina.
\newblock Model of active transport of ions in cardiac cell.
\newblock {\em Journal of Theoretical Biology}, 252(2):247--254, 2008.

\bibitem{Szatmari2020}
D{\'a}vid Szatm{\'a}ri, P{\'e}ter S{\'a}rk{\'a}ny, B{\'e}la Kocsis, Tam{\'a}s
  Nagy, Attila Miseta, Szilvia Bark{\'o}, Be{\'a}ta Longauer, Robert~C.
  Robinson, and Mikl{\'o}s Nyitrai.
\newblock Intracellular ion concentrations and cation-dependent remodelling of
  bacterial mreb assemblies.
\newblock {\em Scientific Reports}, 10(1):12002, Jul 2020.

\bibitem{Sangjin2013}
Sangjin Kim, Erik Broströmer, Dong Xing, Jianshi Jin, Shasha Chong, Hao Ge,
  Siyuan Wang, Chan Gu, Lijiang Yang, Yi~Qin Gao, Xiao dong Su, Yujie Sun, and
  X.~Sunney Xie.
\newblock Probing allostery through {DNA}.
\newblock {\em Science}, 339(6121):816--819, 2013.

\bibitem{Rosenblum2021}
Gabriel Rosenblum, Nadav Elad, Haim Rozenberg, Felix Wiggers, Jakub Jungwirth,
  and Hagen Hofmann.
\newblock Allostery through dna drives phenotype switching.
\newblock {\em Nature Communications}, 12(1):2967, May 2021.

\bibitem{Carlon2023}
Midas Segers, Aderik Voorspoels, Takahiro Sakaue, and Enrico Carlon.
\newblock Mechanisms of dna-mediated allostery.
\newblock {\em Phys. Rev. Lett.}, 131:238402, Dec 2023.

\bibitem{Bosaeus2017}
Niklas Bosaeus, Anna Reymer, Tamás Beke-Somfai, Tom Brown, Masayuki Takahashi,
  Pernilla Wittung-Stafshede, Sandra Rocha, and Bengt Nordén.
\newblock A stretched conformation of dna with a biological role?
\newblock {\em Quarterly Reviews of Biophysics}, 50:e11, 2017.

\bibitem{Safaee2013}
Nozhat Safaee, Anne~M. Noronha, Dmitry Rodionov, Guennadi Kozlov,
  Christopher~J. Wilds, George~M. Sheldrick, and Kalle Gehring.
\newblock Structure of the parallel duplex of poly({A}) {RNA}: Evaluation of a
  50 year-old prediction.
\newblock {\em Angewandte Chemie International Edition}, 52(39):10370--10373,
  2013.

\bibitem{Hartmann2023}
Andreas Hartmann, Koushik Sreenivasa, Mathias Schenkel, Neharika Chamachi,
  Philipp Schake, Georg Krainer, and Michael Schlierf.
\newblock An automated single-molecule {FRET} platform for high-content,
  multiwell plate screening of biomolecular conformations and dynamics.
\newblock {\em Nature Communications}, 14(1):6511, Oct 2023.

\bibitem{Chu2024}
Jiachong Chu, Ayesha Ejaz, Kyle~M. Lin, Madeline~R. Joseph, Aria~E. Coraor,
  D.~Allan Drummond, and Allison~H. Squires.
\newblock Single-molecule fluorescence multiplexing by multi-parameter
  spectroscopic detection of nanostructured {FRET} labels.
\newblock {\em Nature Nanotechnology}, May 2024.

\bibitem{Rissone2024}
Paolo Rissone, Aurélien Severino, Isabel Pastor, and Felix Ritort.
\newblock Universal cold rna phase transitions.
\newblock {\em Proceedings of the National Academy of Sciences},
  121(34):e2408313121, 2024.

\end{thebibliography}

\end{document}